\documentclass[twocolumn,prl,amsmath,amssymb,nopacs]{revtex4}
\usepackage{graphicx}% Include figure files
\usepackage{dcolumn}% Align table columns on decimal point
\usepackage{bm}% bold math
\usepackage{latexsym}
\usepackage{amstext}
\usepackage{amsxtra}
\usepackage{color}

\definecolor{darkblue}{rgb}{0, 0, 0.8}
\usepackage[colorlinks=true, breaklinks=true, linkcolor=darkblue, citecolor=darkblue, urlcolor=darkblue]{hyperref}
\newcommand{\doilink}[2]{\href{http://dx.doi.org/#1}{#2}}

\usepackage{times}

\begin{document}

\title{Realization of a distributed Bragg reflector for propagating guided matter waves}

\author{C. M. Fabre, P. Cheiney, G. L. Gattobigio, F. Vermersch, S. Faure, R. Mathevet, T.~Lahaye and D.~Gu\'ery-Odelin}
\affiliation{Universit\'e de Toulouse, UPS, Laboratoire Collisions Agr\'egats R\'eactivit\'e, IRSAMC; F-31062 Toulouse, France}
\affiliation{CNRS, UMR 5589, F-31062 Toulouse, France}

\date{\today}

\begin{abstract}
We report on the experimental realization of a Bragg reflector for guided matter waves. A Bose-Einstein condensate with controlled velocity distribution impinges onto an attractive optical lattice of finite length and directly probes its band structure. We study the dynamics of the scattering by this potential and compare the results with simple one-dimensional models. We emphasize the importance of taking into account the gaussian envelope of the optical lattice which gives rise to Bragg cavity effects. Our results are a further step towards integrated atom optics setups for quasi-cw matter waves.
\end{abstract}


\maketitle

The interaction of light with structures having a periodic refractive index profile is ubiquitous in photonics. Applications range from simple antireflection coatings to the fabrication of dielectric mirrors with ultra-high reflectivities, used for instance in high-finesse cavities, and to semi-conductor laser technology with the example of VCSELs, and DFB or DBR lasers. In the field of guided optics, fiber Bragg gratings are essential components for the telecommunication industry, as well as for the realization of outcoupling mirrors in high-power fiber lasers. Photonic crystal devices also have a huge range of applications~\cite{saleh2007}.

In matterwave optics and interferometry, interactions of free-space propagating beams or trapped clouds with periodic structures or potentials have been extensively investigated and are commonly used as mirrors and beamsplitters~\cite{cronin2009}. In this paper, we demonstrate, following the proposals of Refs.~\cite{santos1997,santos1998,friedman1998,carusotto2000,lauber2011}, a Bragg reflector for manipulating a guided Bose-Einstein condensate (BEC) propagating in an optical waveguide, i.e. the exact atom-optics counterpart of a photonic fiber Bragg grating. We study the dynamics and the transmission of a probe wavepacket as a function of its incident velocity and of the depth of the optical lattice. As we will develop later on, this quasi 1D configuration exemplifies clearly two textbook features of quantum mechanics: quantum reflection~\cite{cohen1977,shimizu2001,pasquini2004} and band theory~\cite{kittel1953,ashcroft1976}. This article is organized as follows. We first present a simple model to gain some physical insight into the Bragg reflection of a matter wavepacket by a finite-length lattice having a gaussian envelope. Then we describe our experimental implementation and show quantitative agreement between the data and our model. Finally, we discuss numerical simulations that give access to unresolved details in the experiment.

\begin{figure}[b]
\centerline{\includegraphics[width=75mm]{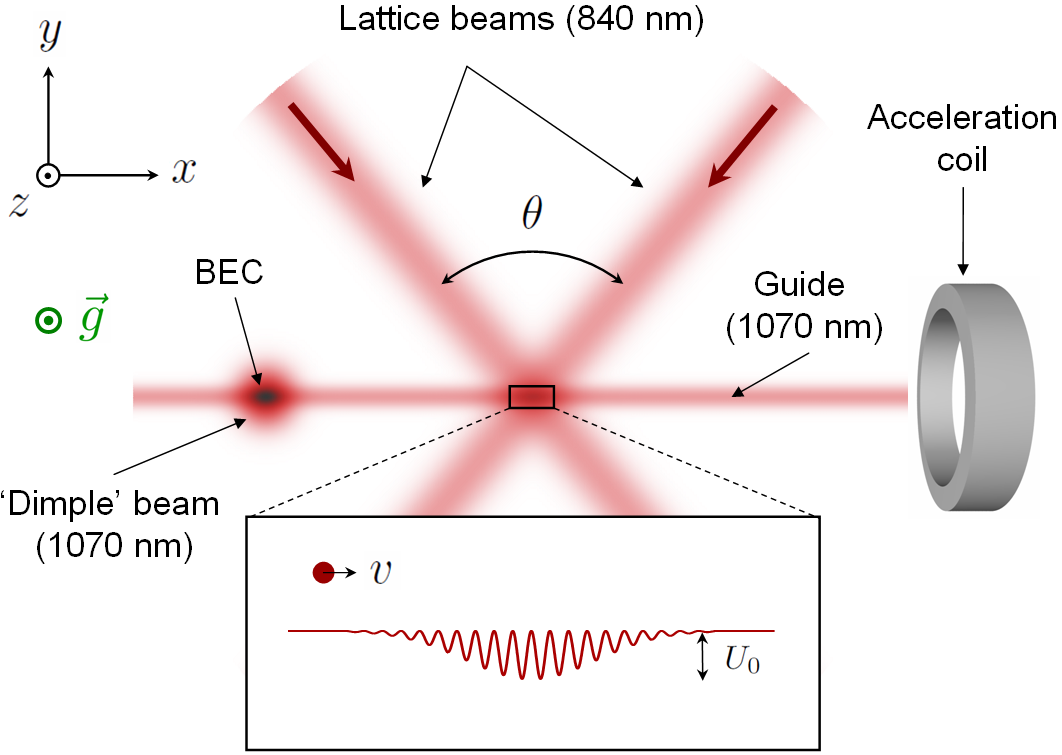}}
\caption{(color online) Schematic view of the experimental setup (not to scale).}
\label{fig1}
\end{figure}

We consider a BEC with given mean velocity $\bar{v}$ and dispersion $\Delta v$ propagating in a horizontal waveguide defining the $x$-axis. At some distance, two intersecting laser beams interfere and create an {\it attractive} quasi-periodic potential of finite length, with lattice spacing $d$ (see Fig.~\ref{fig1}).

The potential experienced by the atoms is modeled by:
\begin{equation}
U(x)=-U_0\exp\left(-\frac{2x^2}{\tilde{w}^2}\right)\sin^2\left(\frac{\pi x}{d}\right),
\label{eq:pot}
\end{equation}
whose depth $U_0$ is proportional to the power of the laser beams. The period $d$ naturally introduces typical scales in velocity $v_{\rm R}=h/(md)$ and energy $E_{\rm R}=mv_{\rm R}^2/2$.

\begin{figure}[t]
\centerline{\includegraphics[width=80mm]{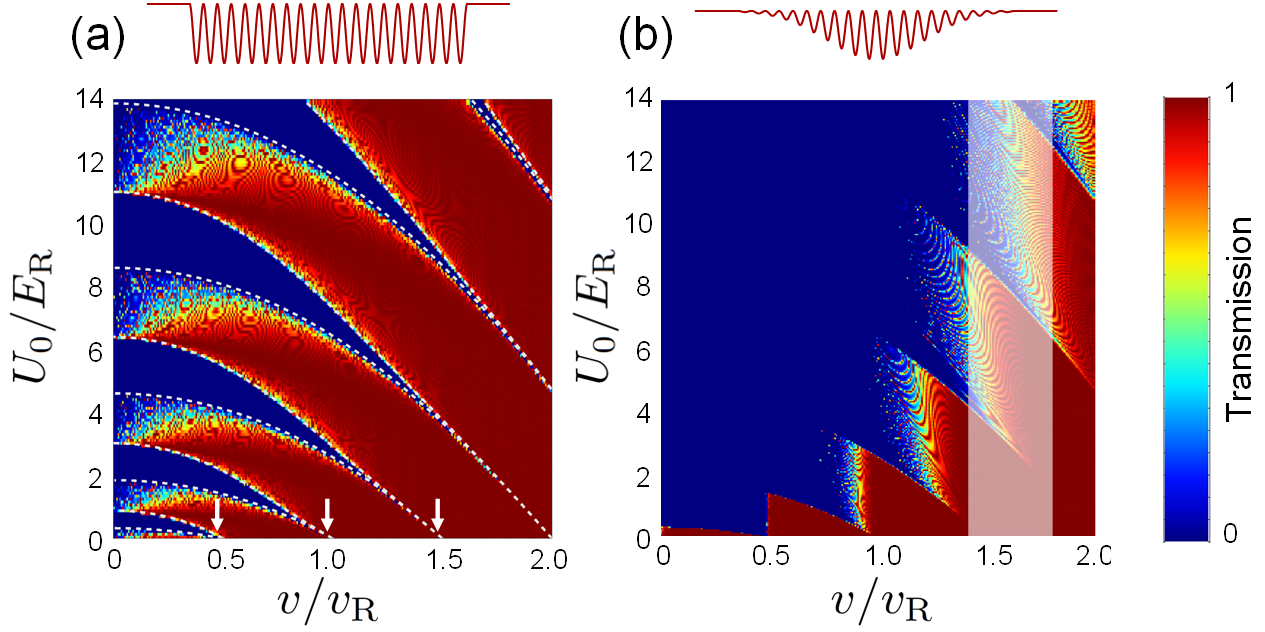}}
\caption{(color online) Transmission coefficient of the lattice as a function of the incident velocity $v$ and of the lattice depth $U_0$. (a): Square-envelope lattice with 800 sites. The white dashed lines are obtained from the Mathieu characteristic functions; white arrows show the velocities for which the Bragg condition is fulfilled (see text). (b):~Lattice with a gaussian envelope ($1/e^2$ radius $\tilde{w}\simeq230d$). The vertical shaded stripe corresponds to the relevant velocity components in the wavepacket used for the measurements shown in Figs.~\ref{fig3} and~\ref{fig4}. The insets on top of panels (a) and (b) illustrate the shape of the lattice enveloppe, but the number of sites is reduced to $N=20$ for clarity.}
\label{fig2}
\end{figure}

We are interested in a wavepacket with finite size and velocity dispersion impinging on a finite-length lattice. Let us consider first the textbook case of an incident plane wave and a square envelope (see e.g.~\cite{sprung1993} for an analytical treatment of the problem). Figure~2(a) shows the transmission coefficient for a lattice of $N=800$~sites, calculated by solving numerically the corresponding stationary Schr\"odinger equation. For a given velocity $v$, one observes that the transmission coefficient essentially switches between 0 and 1 as a function of the depth of the lattice $U_0$. It can be interpreted as follows. At the entrance, the incoming state of energy $E_{\rm i}=mv^2/2$ is projected onto the eigenstates of the lattice. The associated eigenenergies distribute into energy bands~\cite{kittel1953} whose position can be expressed in terms of the characteristic functions of the Mathieu equation~\cite{mclachlan1947,epaps} as depicted by the white dashed lines in Fig.~\ref{fig2}(a). Reflection occurs if $E_{\rm i}$ lies in the gap between two allowed energy bands. Due to the finite length of the lattice the energy bands are not strictly continuous and resolve into $N$ separate states for vanishing incident velocity~\cite{epaps}. Undersampling of the image gives rise to the `foamy' aspect of the low-velocity side of transmission bands. Obviously, the reflection by an \emph{attractive} potential is a purely quantum effect, with no classical counterpart.

A second interesting feature appears in the limit of a vanishing potential depth $U_0$. One still observes reflection but it occurs only for incident velocities of the form $v=nv_{\rm R}/2$ where $n$ is an integer (see the white arrows on Fig.~\ref{fig2}(a)). This corresponds to Bragg's condition $2d\sin\Theta=n\lambda$~\cite{ashcroft1976}, where $\Theta=\pi/2$ for retroreflection, and $\lambda=h/(mv)$ is the incident de Broglie wavelength of the atoms: the reflection amplitude at each lattice site is small but constructive interference between all the reflected waves results in a macroscopic reflected wave building up. For the range of parameters of Fig.~\ref{fig2}(a), quantum reflection by a \emph{single} lattice well occurs only for velocities that are very small as compared to $v_{\rm R}$~\cite{epaps} and thus cannot explain the observed features.

Figure~\ref{fig2}(b) shows the result of the same calculation, but now for the experimentally relevant case of a lattice having a smooth gaussian envelope. One clearly observes a drastic change in the dependence of the transmission coefficient: for a fixed velocity $v$, the transmission is essentially equal to one only \emph{below} a critical value $U_0^{(1)}$ of the lattice depth, and then goes to zero. That critical value corresponds to the smallest one at which total reflection would occur for the square envelope lattice. No resurgence of the transmission is observed if $U_0$ is further increased which gives a `sawtooth' appearance to the boundary between reflection and transmission.

This can be understood as follows. We are in the {\em slowly varying envelope} limit as $\tilde{w}\gg d$. The amplitude of the lattice does not change appreciably over a few sites, and thus one can consider that, locally, the matterwave interacts with a constant-amplitude lattice. When $U_0=U_0^{(1)}$ the reflection condition is met at the center of the lattice i.e. at $x=0$. Then, when $U_0$ increases, there are some locations $\pm x_{\rm refl}$, on both sides of the center, for which $U(\pm x_{\rm refl})=U_0^{(1)}$ and where reflection occurs. This constitutes a Fabry-Perot resonator made of two Bragg mirrors, analog to VCSELs for example. As in optics, transmission exhibits sharp resonances which gives the same foamy aspect as in Fig.~\ref{fig2}(a).

We now come to the experimental realization. Our technique to produce all-optical BECs has been described in details elsewhere~\cite{couvert2008}; in what follows we thus simply recall the major steps. We produce an almost pure $^{87}{\rm Rb}$ condensate containing typically $5\times10^4$ atoms by forced evaporation over 4~s in a crossed optical dipole trap. It is made of two intersecting beams with a wavelength of 1070~nm. A horizontal one, with a waist of $50\;\mu{\rm m}$, to be used later as a guide for the BEC, defines the $\hat{x}$ direction. The second, the `dimple' beam, of waist $150\;\mu{\rm m}$, propagates along the $\hat{x}+\hat{z}$ direction, $\hat{z}$ being the vertical (Fig.~\ref{fig1}). Spin distillation using a magnetic field gradient during evaporation~\cite{couvert2008} results in the BEC being prepared in the state $|F=1,m_F=0\rangle$. We then decrease adiabatically the power in the dimple beam by a factor $\sim 20$ over 80~ms, thus barely keeping a longitudinal confinement for the BEC, before switching it off abruptly to outcouple a wavepacket in the horizontal guide. In this way, we produce a wavepacket with a small longitudinal velocity dispersion. To set the wavepacket in motion, we then switch on a coil, coaxial with the guide, that produces an inhomogeneous magnetic field. Through the quadratic Zeeman effect the wavepacket is accelerated for typically 15~ms to a final mean velocity $\bar{v}$ between 2 and 15 mm$/$s. The residual acceleration of the packet due to stray fields and beams curvature is negligible (we measure an upper bound of $10\;{\rm mm/s^2}$).

Centered 250~$\mu$m downstream from the dimple location $x_0$, the optical lattice is produced at the intersection of two beams with a wavelength $\lambda_{\rm L}=840$~nm (red-detuned with respect to the $^{87}$Rb D1 and D2 lines) and a waist $w=110$~$\mu$m, linearly polarized along $\hat{z}$, crossing at an angle $\theta\simeq 81^\circ$. The lattice detuning is large enough so that spontaneous emission does not play any role on our experimental timescales. The resulting lattice spacing is $d=\lambda_{\rm L}/[2\sin(\theta/2)]\simeq 650$~nm. In a set of preliminary experiments we calibrate the potential depth $U_0$ using Kapitza-Dirac diffraction~\cite{gadway2009,huckans2009}. A BEC is created at the position of the lattice and exposed to the lattice potential for a short time $\tau_{\rm KD}$, typically a few tens of microseconds. The diffraction pattern of the BEC after time of flight as a function of $\tau_{\rm KD}$ is then compared to numerical simulations of the process. A typical 35~mW per beam results in $U_0$ up to $15E_{\rm R}$.

After being launched as described above, the wave packet propagates in the horizontal guide for an adjustable time $t_{\rm prop}$. Then all the lasers are switched off abruptly and the cloud is imaged by absorption after a 10~ms time of flight. This gives access to the spatial density distribution $n(x,t)=|\psi(x,t)|^2$ of the wavepacket with a resolution of about 10~$\mu$m limited by the numerical aperture of our collection lens.

\begin{figure}[t]
\centerline{\includegraphics[width=80mm]{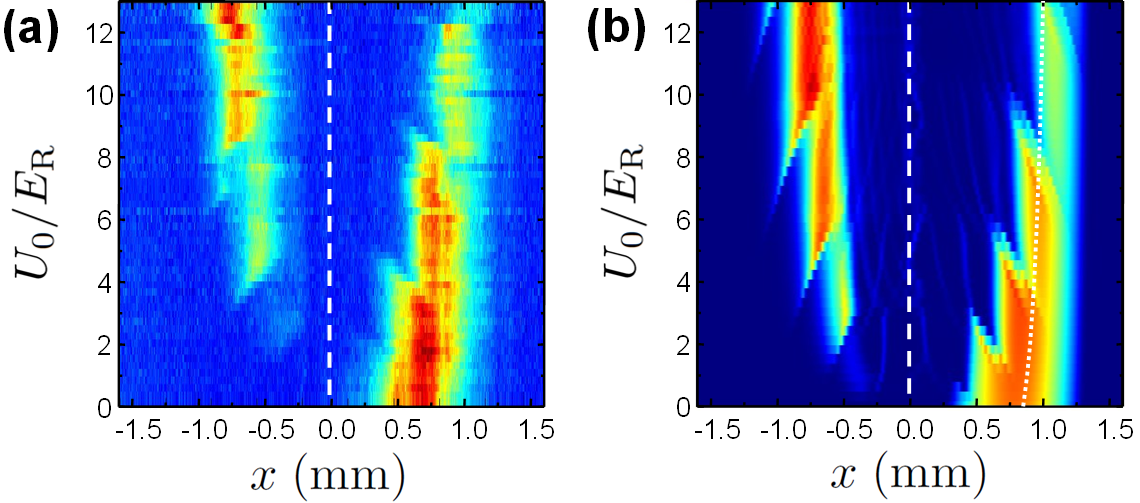}}
\caption{(color online) (a) Measured density distribution of the wavepacket (of initial mean velocity $\bar{v}=11\;{\rm mm/s}$) after a propagation time $t_{\rm prop}=100$~ms, for different lattice depths $U_0$. Each horizontal line is the average of typically eight absorption images integrated along the $\hat{y}$ direction. The vertical white dashed line shows the position of the center of the lattice. (b) Results of the simulation without any adjustable parameters. The finite resolution of the imaging system ($\sim 10\;\mu$m) is included. The dotted line is the expected position of the center of the wavepacket according to classical dynamics (see text).}
\label{fig3}
\end{figure}

In a first set of experiment, the propagation time $t_{\rm prop}=100$~ms is sufficiently long so that the interaction with the lattice is completed. We measured in a separate experiment a mean velocity is $\bar{v}\simeq11\;{\rm mm/s}\simeq1.6\,v_{\rm R}$ and a r.m.s velocity spread $\Delta v\simeq 1.3\;{\rm mm/s}\simeq0.2\,v_{\rm R}$  corresponding to the shaded region of Fig.~\ref{fig2}(b). For each lattice depth $U_0$, an average image is generated from eight individual runs and then integrated along the transverse direction $\hat{y}$. Figure~\ref{fig3}(a) is a stack of 55 such profiles. For sake of comparison, Figure~\ref{fig3}(b) is the result of a numerical simulation of the wavepacket dynamics using the one-dimensional Schr\"odinger equation solved by the split-Fourier method; the initial condition is a gaussian wavepacket with the experimentally measured momentum and position dispersions~\cite{footnote}. There is no adjustable parameter and the overall agreement with experimental data means that our simple 1D model captures most of the physics involved.

Let us concentrate first on the transmitted part of the wavepacket ($x>0$). If there were no lattice, the propagation time $t_{\rm prop}$ is long enough so that the initial size of the wavepacket is negligible with respect to its size after propagation. The spatial distribution of the wavepacket would then be a direct mapping of its initial velocity distribution $f(v)$:
$n(x,t_{\rm prop})\propto f\left[(x-x_0)/t_{\rm prop}\right]$.

Figure~\ref{fig3}(b) compares then directly with Fig.~\ref{fig2}(b) whose horizontal axis is properly scaled. In the background of the shaded area of Fig.~\ref{fig2}(b) representing the wavepacket one can see the transmitted and reflected components. In the presence of the lattice, the reflected part propagates backwards and is located, for the propagation time chosen here, at a symmetrical position. This explains why the transmitted and reflected wavepackets appear like complementary mirrored image of each other. The sawtooth-like boundary, reminiscent of the transmission diagram, is a fingerprint of the band structure inside the lattice. However, the effect of the lattice potential is not limited to the one of the sinusoidal component, responsible for the Bragg reflection described above. The spatially averaged attractive potential also accelerates the wavepacket. The white dotted line on Fig.~\ref{fig3}(b) shows the final position of a classical particle starting with velocity $\bar{v}$ from position $x_0$ and propagating for a time $t_{\rm prop}$, taking into account its acceleration by the spatially averaged lattice potential. The fair agreement with the data indicates that the slight curvature in the position of the wavepacket as a function of $U_0$ simply arises from this classical effect.

\begin{figure}[t]
\begin{center}
\includegraphics[width=75mm]{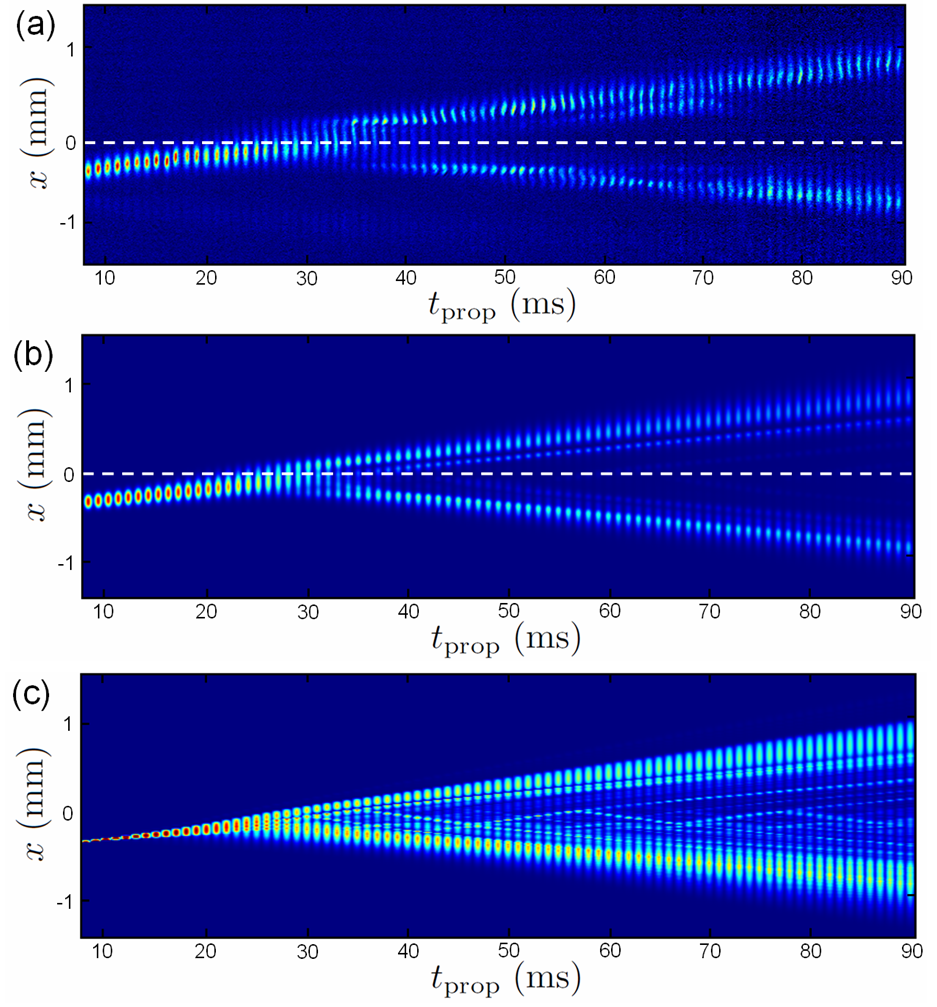}
\end{center}
\caption{(color online) Time sequence showing the scattering of a wavepacket  with mean velocity $\bar{v}\simeq 11\;{\rm mm/s}$ on the optical lattice for $U_0\simeq 11 E_{\rm R}$. The white dashed lines in panels (a) and (b) show the position of the center of the lattice. The time interval between successive images is 1~ms. (a): Experimental data. (b): Simulation, taking into account the finite imaging resolution as well as the time of flight (TOF) period. (c): Same as (b) but without TOF nor reduced resolution; the color scale is nonlinear in order to enhance contrast.}
\label{fig4}
\end{figure}

Beyond studying the asymptotic scattering states, it is also possible to visualise the dynamics of the interaction by varying $t_{\rm prop}$. Figure~\ref{fig4}(a) displays such a time sequence that fairly compares to the numerical simulation depicted in the same conditions in Figure~\ref{fig4}(b). One clearly observes the spreading of the incident wavepacket over the whole lattice for $30\;{\rm ms}\lesssim t_{\rm prop}\lesssim 45\;{\rm ms}$ and its subsequent splitting into a reflected and a transmitted one. Unfortunately, the details of the inner dynamics are washed out by the free expansion of the wavepacket during the time-of-flight sequence and the finite resolution of the imaging system.

Numerical simulations, properly checked against the previous experimental results, are useful here. In Figure~\ref{fig4}(c) we have deliberately suppressed the time-of-flight period and enhanced the optical resolution and the contrast with respect to Fig.~\ref{fig4}(b): one then clearly observes multiple reflections of some components of the wavepacket at symmetric positions
$\pm x_{\rm refl}$, with decreasing amplitude at each bounce. This `cavity-ring-down' behavior explains the formation of structures in the transmitted and reflected wavepackets as observed in Fig.~\ref{fig4}(a) and especially visible as a parallel lower stripe in the transmitted wavepacket for $50\;{\rm ms}\lesssim t_{\rm prop}\lesssim 75\;{\rm ms}$. However, experimentally, observing several bounces is not possible here due to the small number of atoms involved.

\begin{figure}[t]
\centerline{\includegraphics[width=86mm]{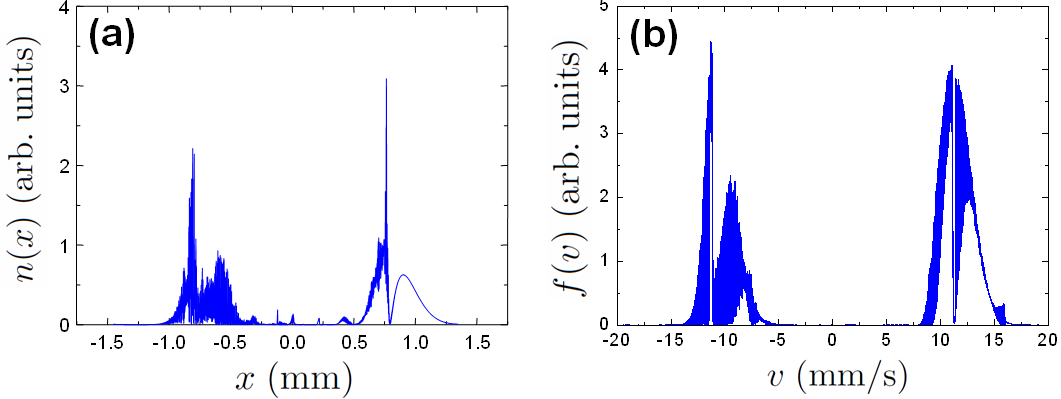}}
\caption{(a) Calculated density  and (b) velocity distribution of a wavepacket of initial mean velocity $\bar{v}=11\;{\rm mm/s}$ after a propagation time $t_{\rm prop}=100$~ms with full resolution. The lattice depth is $U_0=8E_{\rm R}$.}
\label{fig5}
\end{figure}

In the same way, on images such as Fig.~\ref{fig3} the reflected wavepacket appears to be relatively smooth. It is  actually not the case as can be seen on simulations with full resolution (Fig.~\ref{fig5}). The lattice acts as a matterwave interference filter with very narrow features due to the high number of lattice sites (foamy zones of Fig.~\ref{fig2}).

Until now we have used a simple one-dimensional description of the system. However the system is actually far from being one-dimensional, since the transverse quantum of energy $\hbar\omega_\bot\simeq h\times 90$~Hz is much smaller than the typical longitudinal energy scales, typically by two orders of magnitude. Our simple one-dimensional model agrees well with the experimental results as shown above because couplings between longitudinal and transverse degrees of freedom are weak (they are due only to experimental imperfections such as misalignments of the lattice beams with respect to the guide for instance); some transverse excitations can nevertheless be observed on our data (see e.g. the long wavelength dipole oscillations on Fig.~\ref{fig4}(a), especially for $t_{\rm prop}\gtrsim 30$~ms). Stronger couplings would be expected to alter significantly the scattering properties of the structure~\cite{gattobigio2010,gattobigio2011}.

In conclusion, we have studied in details the scattering of a guided matterwave by a finite length optical lattice in the slowly varying envelope limit. The experiments can be interpreted in the framework of a local band structure and the whole lattice can be seen as a Bragg reflector/cavity.

Major improvements are expected with the use of high numerical aperture optics~\cite{sortais2008,henderson2009}. Reducing drastically the length $\tilde{w}$ of the lattice and thus generating a structure consisting of only a few sites, possibly with a shaped envelope, one could tailor almost arbitrarily the matterwave filter response. Moreover, if a controlled frequency offset between the two interfering beams is introduced, one generates a moving lattice. The transmission band of the filter could then be adjusted at will. Such setups would prove useful in measuring for instance the coherence length~\cite{koehl2001} of guided atom lasers~\cite{guerin2006,billy2007,bernard2010}. In a different direction, it would be appealing to study the effect of interatomic interactions~\cite{morsch2006} on the propagation of the wavepacket, with the possible appearance of soliton trains~\cite{carusotto2002}, or atom-blockade effects~\cite{carusotto2001}. This regime could be reached by using much higher transverse frequencies for the guide, in order to enhance the effects of nonlinearities.

We thank I. Carusotto for useful discussions, and acknowledge support from Agence Nationale de la Recherche (GALOP project), R\'egion Midi-Pyr\'en\'ees, and Institut Universitaire de France.

\newpage

\renewcommand\figurename{Fig. S\!}
\setcounter{figure}{0}

\onecolumngrid
\noindent
\begin{center}
\large{\bf Supplementary material}
\end{center}

\vskip0.7cm

In this supplementary material, we provide additional details concerning several points discussed in the main text about the physics of Bragg reflection by a finite-length, square-envelope lattice.
\vskip0.7cm
\textcolor{white}{.}

\twocolumngrid

\vskip0.7cm

\section{Transmission by a semi-infinite, square-envelope lattice and Mathieu characteristic functions}

Here we detail how one can obtain analytically, using the theory of the Mathieu equation, the regions in parameter space $(v,U_0)$ corresponding to total reflection by a semi-infinite lattice.

\begin{figure}[b]
\centerline{\includegraphics[width=75mm]{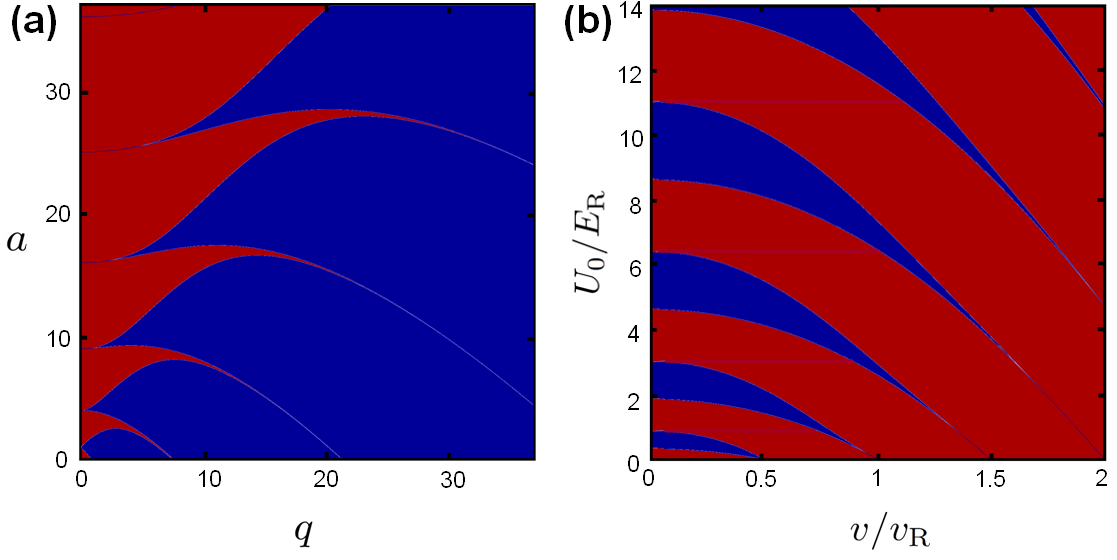}}
\caption{(a): Stability diagram of the Mathieu equation in the $(q,a)$ plane. The stable regions correspond to the red areas. (b) The same diagram as a function of the parameters $(v,U_0)$, see Eq. (\ref{eq:epaps3}).}
\label{fig:mathieu}
\end{figure}

The time-independent Schr\"odinger equation for a particle of mass $m$ and energy $mv^2/2$ evolving in the potential $-U_0\sin^2(\pi x/d)$ reads:
\begin{equation}
\frac{-\hbar^2}{2m}\,\frac{{\rm d}^2\psi}{{\rm d}x^2}-U_0\sin^2\left(\frac{\pi x}{d}\right)\psi(x)=\frac{mv^2}{2}\psi(x).
\label{eq:epaps1}
\end{equation}
Introducing the dimensionless parameter $\tilde{x}=\pi x / d$, Eqn.~(\ref{eq:epaps1}) can be rewritten as
\begin{equation}
\frac{{\rm d}^2\psi}{{\rm d}\tilde{x}^2}+\left[a-2q\cos(2\tilde{x})\right]\,\psi=0,
\label{eq:epaps2}
\end{equation}
which is precisely the definition of the Mathieu equation~\cite{mclachlan1947}, with $a$ and $q$ defined by
\begin{equation}
\left\{
\begin{array}{rcl}
a&=&4\left(v/v_{\rm R}\right)^2+2U_0/E_{\rm R},\\
q&=&U_0/E_{\rm R}.
\end{array}
\right.
\label{eq:epaps3}
\end{equation}

The theory of the Mathieu equation~(\ref{eq:epaps2}) shows that, depending on the parameters $(q,a)$, either the solutions remain bounded whatever the initial conditions (stable regions of the $(q,a)$ plane), or there is at least one solution that grows exponentially when $\tilde{x}\to\infty$~\cite{mclachlan1947}. The stable regions in the $(q,a)$ plane are shown in red in Fig.~S\ref{fig:mathieu}(a). The equation of the boundaries between stable and unstable regions are given by the so-called \emph{Mathieu characteristic functions} $a_n(q)$ and $b_n(q)$, where $n\in\mathbb{N}$.

For the problem of a quantum particle impinging with velocity $v$ on a semi-infinite lattice of depth $U_0$, one easily finds, using the mapping (\ref{eq:epaps3}), the regions in the $(v,U_0)$ parameter space that correspond to unstable solutions of the Mathieu equation. They correspond to situations in which the energy of the incident particle lies in a forbidden band, implying a total reflection by the lattice. These are shown in blue on Fig.~S\ref{fig:mathieu}(b) and agree well with the regions of full reflection by a finite-length lattice shown in the main text on Fig.~2(a).

However, for parameters corresponding to full transmission (in red on Fig.~S\ref{fig:mathieu}(b)), and especially at low velocities, one clearly sees on Fig.~2(a) of the main text that for a finite-length lattice, significant reflections are observed in the regions with a `foamy' aspect. These arise from the finite number of lattice sites, as we shall see in the section below.

\section{Influence of the number of lattice sites for a square-envelope lattice}

\begin{figure*}[t]
\centerline{\includegraphics[width=120mm]{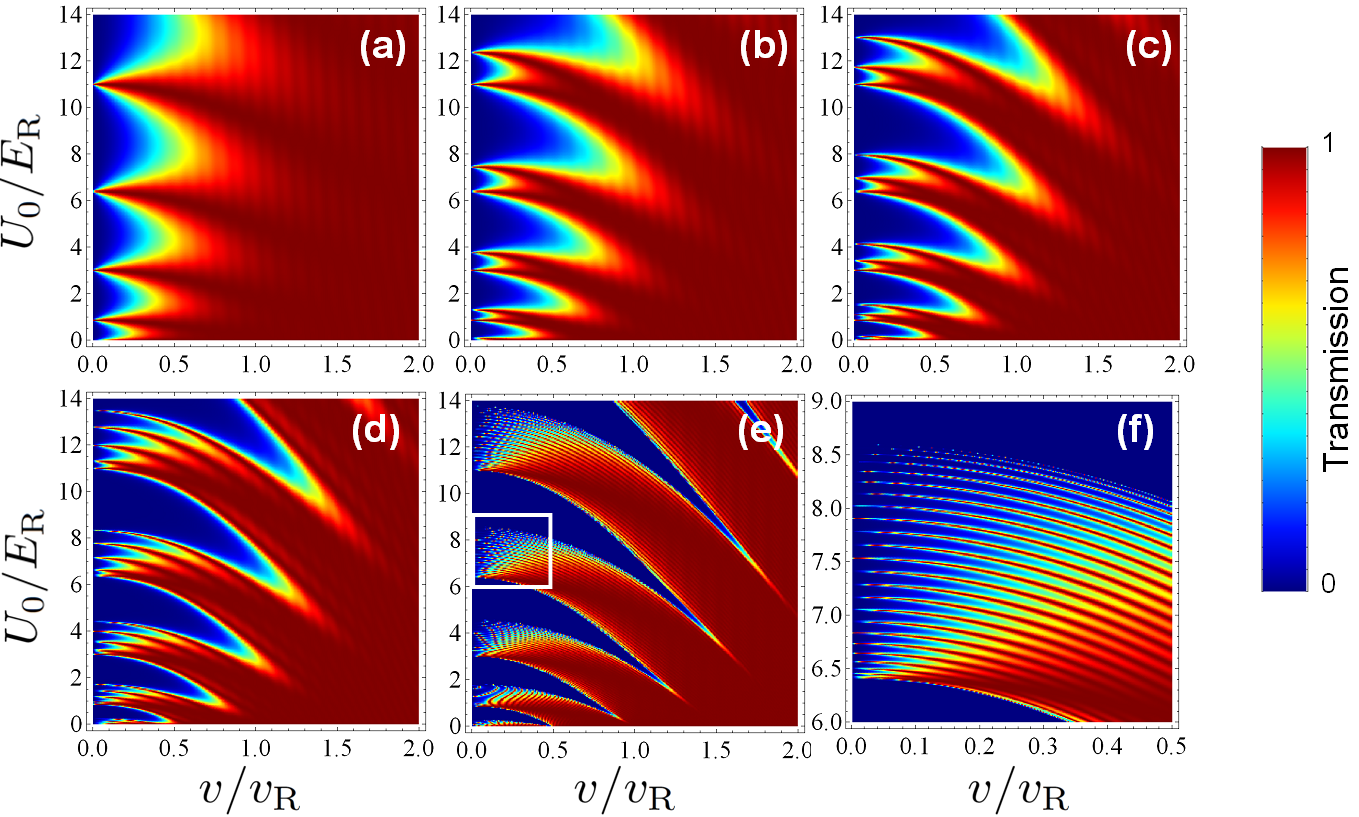}}
\caption{Transmission coefficient of a square-envelope lattice of (a)--(e): $N=1,2,3,5,25$ sites. (f): Magnification of the region in the white rectangle in (e), showing clearly that the allowed ``band'' is actually made of $N$ narrow resonances.}
\label{fig:influence:of:n}
\end{figure*}

The above analytical results are valid for a semi-infinite lattice. For the more experimentally relevant case of a finite-length lattice, we solve numerically the stationary Schr\"odinger equation.
Figure~S\ref{fig:influence:of:n} shows, for increasing numbers of sites $N$ of the lattice, the calculated transmission as a function of $(v,U_0)$.

Figure~S\ref{fig:influence:of:n}(a) corresponds to the transmission of a single sinusoidal site. One clearly observes \emph{quantum reflection} at low velocities (typically, below $0.3\,v_{\rm R}$ for the range of depths $U_0$ studied here), except for discrete values of $U_0$ that correspond to the appearance of a new bound state in the attractive well: in that case instead, full transmission occurs. Note that in order to observe such a quantum reflection by a \emph{single} lattice site with the velocities of $1.6\,v_{\rm R}$ used in the experiments described in the main text, one would need to have a much deeper well (typically several $10^3 E_{\rm R}$).

When increasing the number $N$ of sites, two main effects arise (Fig.~S\ref{fig:influence:of:n}(b)--(e)):
\begin{itemize}
\item At low velocities, the discrete values of $U_0$ corresponding to transmission split into exactly $N$ components, due to the coupling between adjacent sites~\cite{cohen1977}. When the number of sites becomes large, these transmission resonance spread all over the allowed energy bands given by the Mathieu equation as can be seen by comparing Figs.~S\ref{fig:influence:of:n}(e) and S\ref{fig:mathieu}(b). The width of these resonances thus scales as $1/N$ with the number $N$ of sites. For our experimental parameters ($N\sim800$), they are extremely narrow and can be observed only in the numerical simulations (see Fig.~5 of the main text).
\item The regions in the $(v,U_0)$ plane in which total reflection occurs extend towards higher velocities, approaching the ones obtained in the limit $N\to\infty$ in the section above: in that case, the reflection originates from Bragg constructive interference of the waves scattered by multiple sites, and not from quantum reflection on the steep slope of a single attractive potential.
\end{itemize}


\begin{thebibliography}{99}
\bibitem{footnote}
Non-linearities due to atom-atom interactions are negligible for the dilute wavepackets studied here, and thus are not accounted for in the simulation. The only effect of interactions is to set the velocity dispersion of the wavepacket in the initial stages of the expansion in the guide, by conversion of the interaction energy of the BEC into kinetic energy.
\end{thebibliography}
\end{document}